\documentstyle[prl,aps,twocolumn,epsf,floats]{revtex}
\begin{document}
\draft
\wideabs{
\title{OPENING AN ENERGY GAP IN AN ELECTRON DOUBLE LAYER SYSTEM AT
INTEGER FILLING FACTOR IN A TILTED MAGNETIC FIELD}
\author{E.V.~Deviatov, V.S.~Khrapai, A.A.~Shashkin, V.T.~Dolgopolov}
\address{Institute of Solid State Physics, Chernogolovka, Moscow
District 142432, Russia}
\author{F.~Hastreiter, A.~Wixforth}
\address{Ludwig-Maximilians-Universit\"at, Geschwister-Scholl-Platz
1, D-80539 M\"unchen, Germany}
\author{K.L.~Campman, A.C.~Gossard}
\address{Materials Department and Center for Quantized Electronic
Structures, University of California, Santa Barbara, California
93106, USA}
\maketitle

\begin{abstract}
We employ magnetocapacitance measurements to study the spectrum of a
double layer system with gate-voltage-tuned electron density
distributions in tilted magnetic fields. For the dissipative state in
normal magnetic fields at filling factor $\nu=3$ and 4, a parallel
magnetic field component is found to give rise to opening a gap at
the Fermi level. We account for the effect in terms of
parallel-field-caused orthogonality breaking of the Landau wave
functions with different quantum numbers for two subbands.
\end{abstract}
\pacs{PACS numbers: 72.20 My, 73.40 Kp}}

Much interest in electron double layers is attracted by their
many-body properties in a quantizing magnetic field. These include
the fractional quantum Hall effect at filling factor $\nu=1/2$
\cite{chak}, the many-body quantum Hall plateau at $\nu=1$
\cite{murphy}, broken-symmetry states at fractional fillings
\cite{jung}, the canted antiferromagnetic state at $\nu=2$
\cite{zheng}, etc. Still, the single-electron properties of double
layer systems that can be interpreted without appealing exchange and
correlation effects are not less intriguing. A standard double layer
with interlayer distance of about the Bohr radius is a soft
two-subband system if brought into the imbalance regime in which the
electron density distribution is two asymmetric maxima corresponding
to two electron layers. In such a system a small interlayer charge
transfer shifts significantly the Landau level sets' positions;
particularly, the transfer of all electrons in a single quantum level
would lead to a shift as large as the cyclotron energy. In a double
layer system with gate-bias-controllable electron density
distributions at normal magnetic fields, peculiarities were observed
in the Landau level fan chart: at fixed integer filling factor
$\nu>2$ the Landau levels for two electron subbands pin to the Fermi
level over wide regions of a magnetic field, giving rise to a zero
activation energy for the conductivity \cite{davies,dolgop}. The
pinning effect is obviously possible due to the orthogonality of the
Landau level wave functions with different index and, therefore, it
might disappear if the orthogonality were lost for some reason. In
contrast, at $\nu=1$ and 2 the gap was found to be similar to the
symmetric-antisymmetric splitting at balance and having a finite
value for any field. This was explained to be caused by a subband
wave function reconstruction in the growth direction \cite{dolgop}.

Here, we study the electron spectrum of a gate-voltage-tunable double
layer in tilted magnetic fields. We find that for the dissipative
state at filling factor $\nu=3$ and 4 in a normal magnetic field, the
addition of a parallel field component leads to the appearance of a
gap at the Fermi level as indicated by activated conductivity. These
findings are explained in terms of a wave-function
orthogonality-breaking effect caused by parallel magnetic field
component.

\begin{figure}[t]
\centerline{
\epsfxsize=\columnwidth
\epsffile{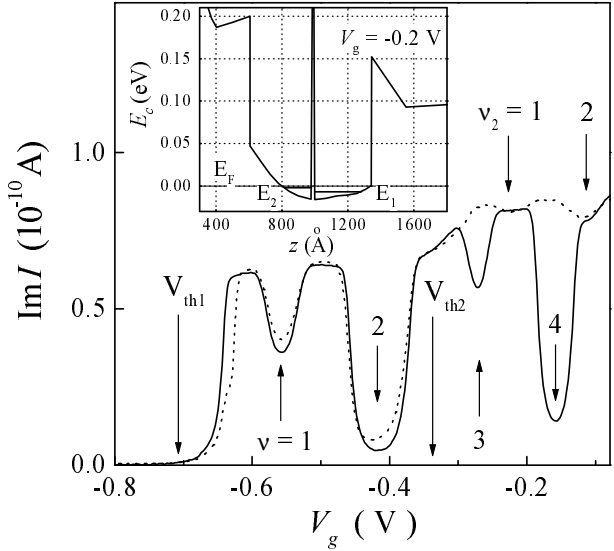}
}
\caption{
Dependence of the imaginary current component on gate
voltage at a frequency of 100~Hz and temperature of 30~mK for a
normal (dashed line) and a tilted by $\Theta=30^\circ$ (solid line)
magnetic field corresponding to the same $B_\perp=3.3$~T. The filling
factors in the whole electron system and in the second subband are
indicated. The threshold voltages are determined from
Fig.~\protect\ref{exp} taking account of insignificant threshold
shifts in different coolings of the sample. Inset: calculated band
diagram of the sample at $V_g=-0.2$~V. Two electron subbands $E_1$
and $E_2$ in the quantum well are filled. The coordinate $z$ is
counted from the gate.
\label{cap}}
\end{figure}

The samples are grown by molecular beam epitaxy on semi-insulating
GaAs substrate. The active layers form a 760~\AA\ wide parabolic
well. In the center of the well a 3 monolayer thick
Al$_x$Ga$_{1-x}$As ($x=0.3$) sheet is grown which serves as a tunnel
barrier between both parts on either side. The symmetrically doped
well is capped by 600~\AA\ AlGaAs and 40~\AA\ GaAs layers. The
symmetric-antisymmetric splitting in the bilayer electron system as
determined from far infrared measurements and model calculations
\cite{hart} is equal to $\Delta_{SAS}=1.3$~meV. The sample has ohmic
contacts (each of them is connected to both electron systems in two
parts of the well) and two gates on the crystal surface with areas
$120\times 120$ and $220\times 120$ $\mu$m$^2$. The gate electrode
enables us to tune the carrier density in the well, which is equal to
$4.2\times 10^{11}$~cm$^{-2}$ at zero gate bias, and simultaneously
measure the capacitance between the gate and the well. For the
capacitance measurements we additionally apply a small ac voltage
$V_{ac}=2.4$~mV at frequencies in the range 3 -- 600~Hz between the
well and the gate and measure both current components as a function
of gate bias $V_g$ in a normal and tilted magnetic fields in the
temperature interval between 30~mK and 1.2~K. An example of the
imaginary current component is depicted in Fig.~\ref{cap}; also shown
in the inset is the calculated behaviour of the conduction band
bottom for our sample.

The employed experimental technique is similar to magnetotransport
measurements in Corbino geometry: in the low frequency limit, the
active component of the current is inversely proportional to the
dissipative conductivity $\sigma_{xx}$ while the imaginary current
component reflects the thermodynamic density of states in a double
layer system. Activation energy at the minima of $\sigma_{xx}$ for
integer $\nu$ is determined from the temperature dependence of the
corresponding peaks in the active current component.

\begin{figure}[t]
\centerline{
\epsfxsize=\columnwidth
\epsffile{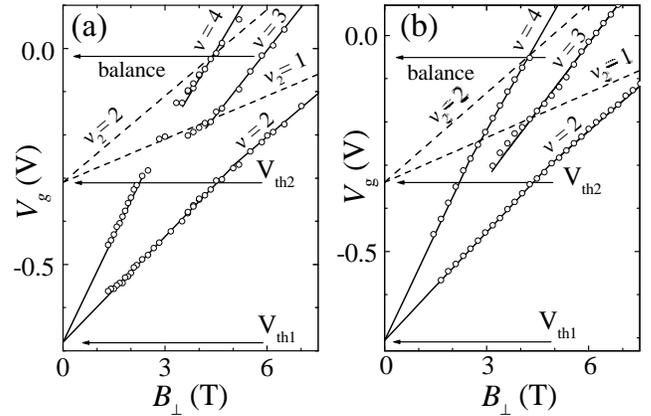}
}
\caption{Positions of the $\sigma_{xx}$ minima at a temperature of
30~mK for different tilt angles: (a) $\Theta=0^\circ$, (b)
$\Theta=30^\circ$. The dashed lines correspond to minima in the
thermodynamic density of states for the second electron
subband.\label{exp}}
\end{figure}

The positions of the $\sigma_{xx}$ minimum for $\nu=2$, 3, and 4 in
the ($B_\perp,V_g$) plane are shown in Fig.~\ref{exp} for both normal
and tilted magnetic field. At the gate voltages $V_{th1} <V_g<
V_{th2}$, at which one subband $E_1$ of the substrate side part of
the well is filled with electrons, the experimental points fall onto
straight lines with slopes defined by capacitance between the gate
and the bottom electron layer. Above $V_{th2}$, where a second
subband $E_2$ collects electrons in the front part of the well, a
minimum in $\sigma_{xx}$ at integer $\nu$ corresponds to a gap in the
spectrum of the bilayer electron system. In this case the slope is
inversely proportional to the capacitance between gate and top
electron layer. Additional minima of the imaginary current component
that are related to the thermodynamic density of states in the second
subband solely are shown in Fig.~\ref{exp} by dashed lines. Hence,
each of the two different kinds of minima forms its own Landau level
fan chart. In the perpendicular magnetic field, wide disruptions of
the fan line at $\nu=4$ and a termination of the line at $\nu=3$
indicate the absence of a minimum in $\sigma_{xx}$ (Fig.~\ref{exp}a).
As mentioned above, this results from a Fermi level pinning of the
Landau levels for two subbands.

Remarkably, switching on a parallel magnetic field is found to
promote the formation of a $\sigma_{xx}$ minimum at integer $\nu>2$,
particularly at $\nu=3$ and 4, see Fig.~\ref{exp}b. This implies that
the parallel magnetic field suppresses the pinning effect, giving
rise to opening a gap at the Fermi level in the double layer system.

\begin{figure}[t]
\centerline{
\epsfxsize=\columnwidth
\epsffile{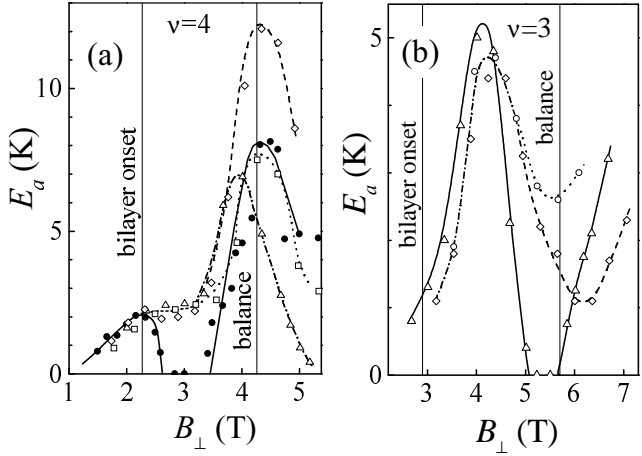}
}
\caption{Change of the activation energy with magnetic field at (a)
$\nu=4$ for $\Theta=0^\circ$ (dots), $\Theta=30^\circ$ (diamonds),
$\Theta=45^\circ$ (squares), $\Theta=60^\circ$ (triangles); and (b)
$\nu=3$ for $\Theta=30^\circ$ (circles), $\Theta=45^\circ$
(diamonds), $\Theta=60^\circ$ (triangles). The lines are guides to
the eye.\label{Ea}}
\end{figure}

Figure~\ref{Ea} represents the behaviour of the activation energy
$E_a$ along the $\nu=3$ and 4 fan lines in Fig.~\ref{exp} for
different tilt angles $\Theta$ of the magnetic field. As seen from
Fig.~\ref{Ea}a, for filling factor $\nu=4$ in the normal field, the
value of $E_a$ is largest both at the bilayer onset $V_{th2}$ and at
balance. In between these it zeroes, which is in agreement with the
disappearance of the minimum of $\sigma_{xx}$ in the magnetic field
range between 2.6 and 3.4~T; in the close vicinity of $B=3$~T, $E_a$
is unmeasurably small but likely finite as can be reconciled with the
observed $\sigma_{xx}$ minimum at the fan crossing point of $\nu=4$
and $\nu_2=1$ (Fig.~\ref{exp}a). In contrast, for tilted magnetic
fields, the activation energy at $\nu=4$ never tends to zero, forming
a plateau, instead (Fig.~\ref{Ea}a).

For $\nu=3$ the parallel field effects are basically similar to the
case of $\nu=4$ with one noteworthy distinction. Near the balance
point, the activation energy in a tilted magnetic field exhibits a
minimum that deepens with increasing tilt angle, see Fig.~\ref{Ea}b.
This minimum is likely to be of many-body origin: at sufficiently
large $\Theta$ it is accompanied by a splitting of the $\nu=3$ fan
line, which is very similar to the behaviour of the double layer at
$\nu=2$ discussed as manifestation of the canted antiferromagnetic
phase \cite{zheng}. This effect will be considered in detail
elsewhere.

We relate the appearance of a gap at integer $\nu>2$ in the
unbalanced double layer at tilted magnetic fields to orthogonality
breaking of the Landau wave functions with different quantum numbers
for two subbands. Indeed, the interlayer tunneling should occur with
in-plane momentum conservation so that in a tilted magnetic field it
is accompanied with an in-plane shift \cite{bet} of the center of the
Landau wave function by an amount $d_0\tan\Theta$, where $d_0$ is the
distance between the centers of mass for electron density
distributions in two lowest subbands. Apparently, the so-shifted
Landau wave functions with different quantum numbers for two subbands
become overlapped. In this case the above mentioned pinning effect at
integer $\nu>2$ cannot occur any more. Instead, as will be discussed
below, the wave functions get reconstructed, which is accompanied by
the levels' splitting.

We calculate the single-particle spectrum in a tilted magnetic field
in self-consistent Hartree approximation without taking into account
the spin splitting (supposing small $g$ factor) as well as the
exchange and correlation energy. The intersubband charge transfer
when switching on the magnetic field is a perturbation potential in
the problem that mixes the wave functions for two subbands. Account
is taken of a shift of the subband bottoms due to a parallel
component of the magnetic field, and the value of gap at the Fermi
level is determined in the first order of perturbation theory in a
similar way to the $\nu=1$ and 2 case at normal magnetic fields of
Ref.~\cite{dolgop}.

\begin{figure}[t]
\centerline{
\epsfxsize=\columnwidth
\epsffile{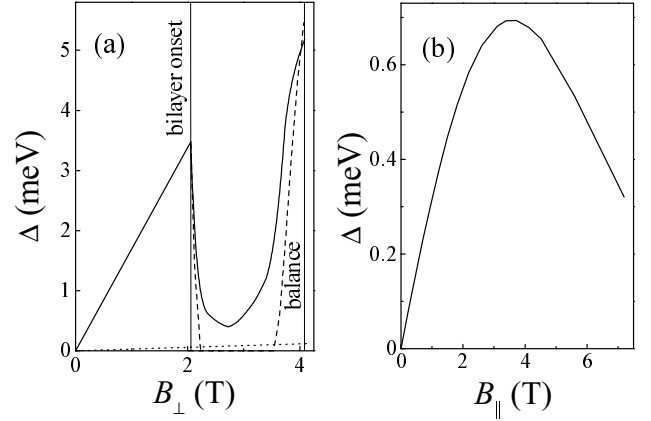}
}
\caption{The calculated gap at $\nu=4$ as a function of magnetic
field for (a) fixed tilt angle $\Theta=0^\circ$ (dashed line) and
$\Theta=30^\circ$ (solid line); and (b) fixed $B_\perp=2.6$~T. Also
shown by a dotted line is the corresponding Zeeman splitting for
$\Theta=30^\circ$.\label{calc}}
\end{figure}

The magnetic field dependence of the calculated gap $\Delta$ for
filling factor $\nu=4$ is displayed in Fig.~\ref{calc}. At fixed tilt
angle the calculation reproduces well the observed behaviour of the
gap along the $\nu=4$ fan line ({\em cf.} Figs.~\ref{Ea}a and
\ref{calc}a). The quantitative difference between the gap values can
be attributed to the finite width of the Landau levels which is
disregarded in calculation.

The gap $\Delta$ as a function of parallel magnetic field component
$B_\parallel$ at a fixed value of $B_\perp=2.6$~T is depicted in
Fig.~\ref{calc}b. It reaches a maximum at $B_\parallel=3.5$~T and
then drops with further increasing field $B_\parallel$. It is clear
that $\Delta(B_\parallel)$ reflects the dependence of the overlap of
the Landau wave functions with different quantum numbers on their
in-plane shift $d_0\tan\Theta$: while at sufficiently small shifts
the overlap rises with shift, at large shifts the overlap is sure to
vanish, restoring the wave function orthogonality.

The above explanation holds for filling factor $\nu=3$ as well. We
note that, in a normal magnetic field, the $\nu=3$ gap for our case
is of spin origin since the expected spin splitting is smaller than
$\Delta_{SAS}$ \cite{dolgop,dens}. Therefore, it can increase with
$B_\parallel$ for trivial reasons. The point of importance is that
the Landau wave function orthogonality has to be lost for the gap to
open.

In summary, we have performed magnetocapacitance measurements on a
double layer system with gate-voltage-controlled electron density
distributions in tilted magnetic fields. It has been found that, for
the dissipative state in normal magnetic fields at filling factor
$\nu=3$ and 4, a parallel magnetic field component leads to opening a
gap at the Fermi level. We attribute the origin of the effect to
orthogonality breaking of the Landau wave functions with different
quantum numbers for two subbands as caused by parallel magnetic
field. The calculated behaviour of the gap is consistent with the
experimental data.

We are thankful to S.V.~Iordanskii for valuable discussions. This
work was supported in part by the Deutsche Forschungsgemeinschaft
DFG, the AFOSR under Grant No.~F49620-94-1-0158, the Russian
Foundation for Basic Research under Grants No.~00-02-17294 and
No.~98-02-16632, the Programme "Nanostructures" from the Russian
Ministry of Sciences under Grant No.~97-1024, and INTAS under Grant
No.~97-31980. The Munich - Santa Barbara collaboration has also been
supported by a joint NSF-European Grant and the Max-Planck research
award.

\end{document}